\documentclass[10pt,a4paper,pre,twocolumn,showpacs,superscriptaddress,bibnotes]{revtex4}
\usepackage{hyperref}
\usepackage{graphicx}
\usepackage{amsmath}
\usepackage{amssymb}
\usepackage{stmaryrd}
\usepackage{color}

\newcommand{\dd}{d}
\usepackage{color}
\newcommand{\yaming}[1]{\textcolor{black}{#1}}

\begin{document}


\title{First-passage time of Brownian motion with dry friction}
\author{Yaming Chen}
\email{yaming.chen@qmul.ac.uk}
\affiliation{School of Mathematical Sciences, Queen Mary University of London, London E1 4NS, United Kingdom}

\author{Wolfram Just}
\email{w.just@qmul.ac.uk}
\affiliation{School of Mathematical Sciences, Queen Mary University of London, London E1 4NS, United Kingdom}

\date{March 18, 2014}

\begin{abstract}
We provide an analytic solution to the first-passage time (FPT) problem of a piecewise-smooth stochastic model, namely Brownian motion with dry friction, using two different but closely related approaches which are based on eigenfunction decompositions on the one hand and on the backward Kolmogorov equation on the other. For the simple case containing only dry friction, a phase transition phenomenon in the spectrum is found which relates to the position of the exit point, and
which affects the tail of the FPT distribution. For the model containing as well a driving force and viscous friction
the impact of the corresponding stick-slip transition and of the transition to ballistic exit is evaluated quantitatively. The proposed model is one of the very few cases where FPT properties are accessible by analytical means.\end{abstract}

\maketitle

\section{Introduction}\label{sect1}

The study of first-passage time (FPT) problems has a very long tradition with its roots in the first half of the last
century by the seminal study of Kramers on chemical kinetics \cite{Kram_P40} (see also Ref.~\cite{HaTaBo_RMP90}
for an excellent review). While FPT problems originated in physical chemistry concepts of this type have turned out
to be relevant in diverse disciplines, like mathematical finance \cite{Linetsky2004ou}, biological modelling
\cite{TuckwellWanRospars2002ou}, complex media \cite{CondaminBenichou2007FPT}, and others.
In an abstract setting the FPT is defined as the time when a stochastic process, often governed by a
stochastic differential equation (SDE), exits a given region for the first time. Beyond the classical setup
problems of this type are relevant in different subjects. Renewed interest in FPT problems has been triggered by studies to characterize large deviation properties, extreme events,
and nonequilibrium processes in many particle systems (see, e.g., Refs.~\cite{Redner2001FPT,BrayMajumdar2013Persistence}).
FPT problems are normally nontrivial to solve and a deeper analytical understanding of FPT properties, e.g., the
dependence on parameters of the system is often hampered by the lack of analytically tractable model systems.
\yaming{There exists a vast literature about this topic, whereby applications often require the application
of numerical tools. Various simple model systems can be handled by analytical means. Among those are the pure diffusion
process \cite{Majumdar2005Brownian}, the Brownian motion with constant drift
\cite{KearneyMajumdar2005FirstPassage}, to some extent the Ornstein-Uhlenbeck process
\cite{Siegert1951FPT,AliliPatiePedersen2005ou} and Bessel processes \cite{GoingYor2003,DeBlassieSmits2007}}.
It is one aim of the present study to provide analytic insight
into a FPT problem which has some relevance for the phenomenological description of friction processes often used
in the engineering context.

Dynamical systems with discontinuities are frequently used for the phenomenological modelling in engineering science.  The impact of such discontinuities on dynamical behavior has attracted recently considerable attention from the general dynamical systems point of view (see, e.g., Ref.~\cite{makarenkov2012}). While the general mathematical theory as well as the theory of corresponding stochastic models is still incomplete, models of such a type have been used successfully in the engineering context for decades. The most prominent examples are dry friction processes, which themselves are not fully understood from the microscopic point of view (see, e.g., Ref~\cite{VaMaUrZaTo_RMP13}). Here we want to go beyond the deterministic dynamical systems setup and intend to study the interrelation between noise and discontinuities, in particular, with regards to FPT problems. We aim at an analytic investigation of a simple piecewise-smooth stochastic model. While some exact results for the propagator of a few simple  piecewise-constant or piecewise-linear SDEs have been known (see for instance Refs.~\cite{karatzas1984,TouchetteStraeten2010Brownian,TouchetteThomas2012Brownian,simpson2012}),
exact results for the FPT problems of piecewise-smooth SDEs are to the best of our knowledge not available in the literature.

To investigate the effect of discontinuities on a FPT problem we take as a motivation Brownian motion with dry (also called solid or Coulomb) friction \cite{Gennes2005dryfriction,Hayakawa2005Langevin}. We consider for our analytic investigations a paradigmatic model system, the phenomenological description of a particle subjected to dry and viscous friction, noise, and a static driving force, resulting in a piecewise-linear SDE
\begin{equation}\label{aa}
\dot{v}(t)=-\mu\sigma(v(t))-\gamma v(t)+b+\sqrt{D}\xi(t).
\end{equation}
Here $ \sigma(v) $, denoting the sign of $v$, represents the dry friction force with coefficient $ \mu>0 $, $ \gamma \geq 0 $ denotes the viscous friction coefficient, $ b $ is a constant biased force and $ D>0 $ is the strength of the Gaussian white noise $ \xi(t) $ characterized by
\begin{equation}\label{ab}
\langle \xi(t) \rangle=0,\qquad \langle\xi(t)\xi(t')\rangle=2\delta(t-t').
\end{equation}
The notation $ \langle\cdots \rangle $ stands for \yaming{the average over all possible realizations} of the noise.
Physically, this model describes the velocity of a solid object of unit mass sliding over an inclined surface with dry and viscous friction. Since the motion of two solid objects over each other is a ubiquitous problem in nature, the dry friction model (\ref{aa}) is important to understand the underlying dynamics of the motion. Mathematically, Eq.~(\ref{aa}) is a piecewise-linear SDE, which allows us to obtain analytic results. For instance, expressions for the propagator can be derived analytically by using spectral decomposition methods \cite{TouchetteStraeten2010Brownian} or Laplace transforms \cite{TouchetteThomas2012Brownian}. \yaming{In particular the propagator of the pure dry friction case (also called Brownian motion with two-valued drift, i.e., Eq.~(\ref{aa}) with $ \gamma=b=0 $) is available in closed analytic form (see, e.g., Refs.~\cite{karatzas1984,KaratzasShreve1991,TouchetteStraeten2010Brownian}).} The weak-noise limit of the model (\ref{aa}) has also been studied in detail by using a path integral approach \cite{BauleCohenTouchette2010path,BauleTouchetteCohen2011path,ChenBauleTouchetteJust2013Weaknoise}. As a piecewise-smooth SDE \cite{ChenBauleTouchetteJust2013Weaknoise,simpson2012}, Eq.~(\ref{aa}) shows many interesting features such as stick-slip transitions \cite{BauleCohenTouchette2010path,BauleTouchetteCohen2011path} and a noise-dependent decay of correlation functions \cite{TouchetteStraeten2010Brownian}. Some of these features have also been shown experimentally in Refs.~\cite{ChaudhuryMettu2008Brownian,GoohpattaderMettu2009Experimental,
Gnoli2013,GoohpattaderChaudhury2010Diffusive}.

Hence, for such a paradigmatic model it is obvious to have a closer look at the corresponding FPT problem, which is the purpose of this paper. In our investigation we consider the exit from a semi-infinite escape interval $ (a,\infty) $. We can confine the analysis to negative exit points, i.e., $ a<0 $. Otherwise, for  $ a\geq 0 $ the discontinuity at $v=0$ will not enter the
FPT problem and we are left with the well known FPT problem of Brownian motion with constant drift ($ \gamma=0 $)  \cite{KearneyMajumdar2005FirstPassage}
or the Ornstein-Uhlenbeck process ($ \gamma\neq 0 $) \cite{WangUhlenbeck1945Brownian}, respectively.

We address the FPT problem for Eq.~(\ref{aa}) by solving a corresponding Fokker-Planck equation via a spectral
decomposition method on the one hand, and by solving a corresponding backward Kolmogorov equation
on the other (see, e.g., Refs.~\cite{Risken1989FPE,Gardiner1990StoMeth}). To keep the presentation self-contained these two methods will be briefly revisited in Section \ref{sect2}. In Section \ref{sect3}, we apply these methods to solve the seemly trivial case without viscous friction ($ \gamma=0 $) and without bias ($ b=0 $), i.e., the pure dry friction case. This simple example already shows a phase transition phenomenon in the spectrum which is related to the position of the exit point. Thereafter,  in Section \ref{sect4} the distribution of the FPT is derived for the model including viscous friction and external force. Here the focus will be on the stick-slip transition and a transition to ballistic exit.
Results are summarized in Section \ref{sect5}.

\section{Remarks on the FPT problem}
\label{sect2}

The approach to FPT problems is well documented in the literature, and suitable expositions can be found in standard textbooks, e.g., Ref.~\cite{Risken1989FPE}. Here we just summarize the essential ideas not only for the convenience of the reader but also to address the few technical issues related to piecewise-smooth drifts. We will focus on the Langevin equation
\begin{equation}\label{ba}
\dot{v}=-\Phi'(v)+\xi(t),
\end{equation}
where the potential $\Phi(v)$ is smooth everywhere apart from $v=0$ and its derivative may have a discontinuity. In particular we will compare and contrast two different but closely related approaches based on eigenfunction decompositions on the one hand and on the backward Kolmogorov equation on the other.

\subsection{Spectral decomposition}

If one considers the stochastic dynamics according to Eq.~(\ref{ba}) on the interval $ (a,\infty) $ it is well known that  the corresponding distribution of the FPT for orbits starting at $v(0)=v_0 \in (a,\infty)$ is given by (see Ref.~\cite{Risken1989FPE})
\begin{equation}\label{bb}
f(T,v_0)=-\frac{\partial }{\partial T}\int_{a}^{\infty}p(v,T|v_0,0)\dd v,
\end{equation}
where the propagator $p(v,t|v_0,0)$ satisfies the corresponding Fokker-Planck equation
\begin{equation}\label{bc}
\frac{\partial }{\partial t}p(v,t|v_0,0)=\frac{\partial}{\partial v}[\Phi'(v) p(v,t|v_0,0)]+\frac{\partial^2}{\partial v^2}p(v,t|v_0,0)
\end{equation}
with an initial condition
\begin{equation} \label{bd}
p(v,0|v_0,0)=\delta(v-v_0),
\end{equation}
an absorbing boundary condition at the left interval endpoint
\begin{equation} \label{be}
p(a,t|v_0,0)=0,
\end{equation}
and a reflecting boundary, i.e., a vanishing probability current at infinity.
To get the solution $ p(v,t|v_0,0) $ we follow a spectral decomposition method
for piecewise-smooth systems used, e.g., in  Ref.~\cite{TouchetteStraeten2010Brownian},
and first solve the associated eigenvalue problem of Eqs.~(\ref{bc})--(\ref{be})
\begin{equation} \label{bf}
-\Lambda u_{\Lambda}(v)=[\Phi'(v)u_{\Lambda}(v)]'+u_{\Lambda}''(v)
\end{equation}
with the (formal) boundary conditions
\begin{equation}\label{bf0}
u_{\Lambda}(a)=0, \qquad
\left.[\Phi'(v) u_\Lambda(v)+u'_\Lambda(v)]\right|_{v\rightarrow \infty}=0 .
\end{equation}
Since we are here concerned with the piecewise-smooth potential $ \Phi(v) $,
we have to solve Eq.~(\ref{bf}) on the two domains $ v>0 $ and $ v<0 $, respectively, and have to apply suitable matching conditions, i.e.,
\begin{equation}
u_{\Lambda}(0-)=u_{\Lambda}(0+) \label{bi}
\end{equation}
coming from the continuity of the eigenfunction and
\begin{equation}
\Phi'(0-)u_{\Lambda}(0-)+u'_{\Lambda}(0-)
=\Phi'(0+)u_{\Lambda}(0+)+u'_{\Lambda}(0+) \label{bj}
\end{equation}
from the continuity of the probability current in Eq.~(\ref{bc}). As in the standard case of Fokker-Planck equations with reflecting boundary conditions the eigenfunctions of the Fokker-Planck operator and the eigenfunctions of the formally adjoint problem are related to each other by an exponential factor containing the potential $\Phi(v)$. Furthermore, both types of eigenfunctions are mutually orthogonal sets and thus result in the orthogonality relations
\begin{eqnarray}
&&\int_{a}^{\infty}u_{\Lambda_m}(v)u_{\Lambda_n}(v)e^{\Phi(v) }\dd v= Z_{\Lambda_n}\delta_{mn},\label{bh}
\\
&&\int_{a}^{\infty}u_{\Lambda}(v)u_{\Lambda'}(v)e^{\Phi(v) }\dd v= Z_\Lambda\delta(\Lambda-\Lambda'), \label{bh0}
\end{eqnarray}
depending on whether the eigenvalue is contained in the discrete or the continuous part of the spectrum. These conditions implicitly take the reflecting boundary at infinity into account. Furthermore, it is worth  mentioning that the reasoning for Fokker-Planck equations with reflecting boundary conditions can be also applied to map the eigenvalue problem to a formally Hermitian positive operator (see Refs.~\cite{Risken1989FPE,Gardiner1990StoMeth}). Thus, all eigenvalues are positive, in particular they are real. Finally, the solution of Eq.~(\ref{bc}) is given by (see, e.g., Ref.~\cite{HoLe_84} for an accessible account on the completeness of the spectrum)
\begin{eqnarray}
p(v,t|v_0,0)&=&e^{\Phi(v_0)}\bigg(\sum_{n} u_{\Lambda_n}(v_0)u_{\Lambda_n}(v)e^{-\Lambda_n t }/Z_{\Lambda_n}\nonumber\\
&&+\int u_{\Lambda}(v_0)u_{\Lambda}(v)e^{-\Lambda t}/Z_\Lambda\dd \Lambda\bigg),\label{bg}
\end{eqnarray}
where the sum is taken over the discrete eigenvalues and the integral is taken over the continuous part of the spectrum. The normalization factors
$ Z_{\Lambda_n} $ and $ Z_{\Lambda} $ are determined by Eqs.~(\ref{bh}) and
(\ref{bh0}), respectively.

\subsection{Backward Kolmogorov equation}

The propagator $p(v,t|v_0,0)$, which determines the FPT
distribution (\ref{bb}), obeys the backward Kolmogorov equation
\cite{Gardiner1990StoMeth} with absorbing boundary condition at $v_0=a$ and
reflecting boundary condition at infinity. Hence the FPT
distribution obeys the backward Kolmogorov equation as well, i.e.,
\begin{equation}\label{bfa}
\frac{\partial }{\partial T}f(T,v_0)=-\Phi'(v_0)
\frac{\partial}{\partial v_0}f(T,v_0)
+\frac{\partial^2}{\partial v_0^2} f(T,v_0)
\end{equation}
with initial condition
\begin{equation}\label{bfb}
f(0,v_0)=0\quad \mbox{for } v_0>a .
\end{equation}
The two boundary conditions, i.e., Eq.~(\ref{be}) and vanishing probability current at infinity, translate into
\begin{equation}\label{bfc}
f(T,v_0\rightarrow a)=\delta(T)
\end{equation}
at the left interval endpoint, and into
\begin{equation}\label{bfd}
\frac{\partial}{\partial v_0}f(T,v_0\rightarrow \infty)=0
\end{equation}
at infinity. If we use the Laplace transform
\begin{equation}\label{bl}
\tilde{f}(s,v_0)=\int_{0}^{\infty}f(T,v_0)e^{-sT}\dd T,
\end{equation}
the partial differential equation (\ref{bfa}) turns into
the ordinary boundary value problem
\begin{equation}\label{bq}
\frac{\partial^2 }{\partial v_0^2}\tilde{f}(s,v_0)-\Phi'(v_0)\frac{\partial }{\partial v_0}\tilde{f}(s,v_0)-s \tilde{f}(s,v_0)=0,
\end{equation}
where Eq.~(\ref{bfc}) obviously results in
\begin{equation}\label{br}
\tilde{f}(s,v_0\rightarrow a)=1.
\end{equation}
As for the other boundary condition we observe that the Laplace
transform (\ref{bl}) converges uniformly in $v_0$ for $s$
being in the right half plane, as the integral converges uniformly at
$s=0$. Hence Eq.~(\ref{bfd}) yields
\begin{equation}\label{bra}
\frac{\partial}{\partial v_0}
\tilde{f}(s,v_0\rightarrow \infty)=0\quad \mbox{for } \mbox{Re}(s)>0.
\end{equation}
Intuitively the two boundary conditions (\ref{br}) and (\ref{bra})
take care of the fact that
on the one hand the FPT is $\delta$-distributed in the
limit
$ v_0\rightarrow a $ and that on the other hand the particle cannot exit the
given region $ (a,\infty) $ at infinity. In addition,
Eq.~(\ref{bq}) should be solved for $ v_0>0 $ and $ v_0<0 $ separately with
matching conditions at $ v_0=0 $, i.e.,
\begin{equation}\label{bs}
\tilde{f}(s,0-)=\tilde{f}(s,0+), \qquad \frac{\partial }{\partial v_0}\tilde{f}(s,0-)=\frac{\partial }{\partial v_0}\tilde{f}(s,0+),
\end{equation}
where the first condition follows from the solution $ \tilde{f}(s,v_0) $
being continuous at $ v_0=0 $ and the second one is derived by integrating
Eq.~(\ref{bq}) across $ v_0=0 $.

The approach via the backward Kolmogorov equation enables us to
obtain the Laplace transform of the FPT distribution in closed analytic
form. Even though it may not be possible to perform the inverse
transform by analytical means to compute $f(T,v_0)$, by taking derivatives the moments of the FPT,
$\langle T^n\rangle$, are then easily evaluated as
\begin{equation}\label{bt}
\langle T^n \rangle=(-1)^n\left. \frac{\partial^n }{\partial s^n}\tilde{f}(s,v_0)
\right|_{s=0} \quad\mbox{for } n=1,2,3,\dots
\end{equation}

\section{The inviscid case}
\label{sect3}

Let us first consider the seemingly trivial case without viscous friction ($\gamma=0$)
and without any external bias ($b=0$), i.e., a particle which is
only exposed to dry friction with a piecewise-constant drift term.
We consider this simplest case as it already shows, somehow counterintuitively,
the main phase transition behavior in the FPT distribution. As a by-product
we can also illustrate all the analytical tools in a very transparent
setup.

If we consider Eq.~(\ref{aa}) for $\gamma=b=0$ we can specialize to the
choice $ \mu=D=1 $ without loss of generality. Other nonvanishing
values are covered by the appropriate rescaling
\begin{equation}\label{ca1}
x=\mu v/D, \qquad \tau=\mu^2 t/D.
\end{equation}
Hence, in this case Eq.~(\ref{aa}) can be written in the form (\ref{ba}) with
\begin{equation}\label{ca0}
\Phi(v)=|v|.
\end{equation}
The corresponding eigenvalue problem (\ref{bf}) consists of a discrete
eigenvalue for $ \Lambda<1/4 $ and a continuous spectrum for $ \Lambda>1/4 $
(cf.\ also Ref.~\cite{Wong1964}).
The details of the derivation are summarized in Appendix \ref{appA} for the convenience
of the reader.

\begin{figure*}
\begin{center}
\includegraphics[scale=0.8]{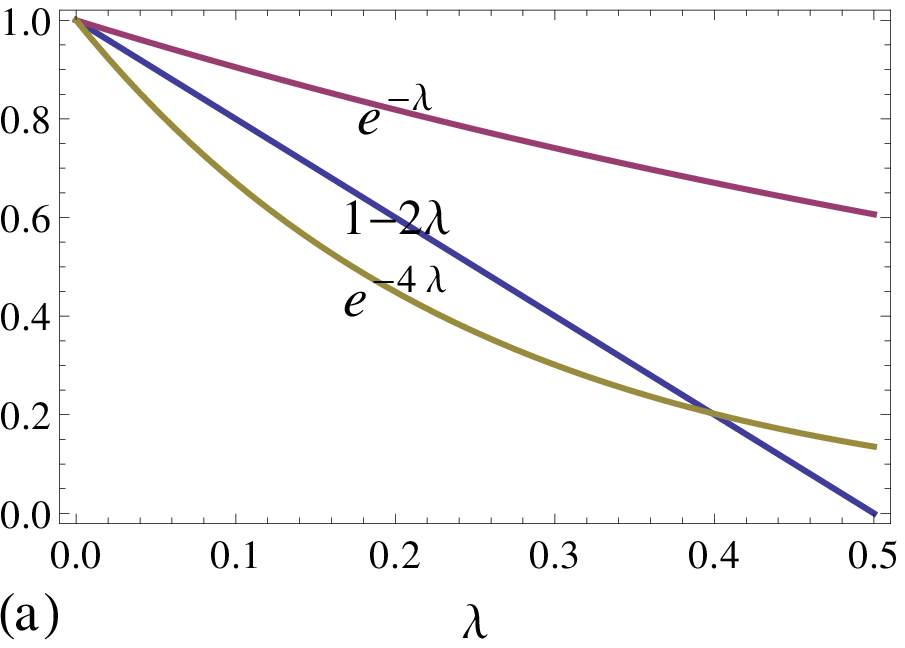}
\includegraphics[scale=0.9]{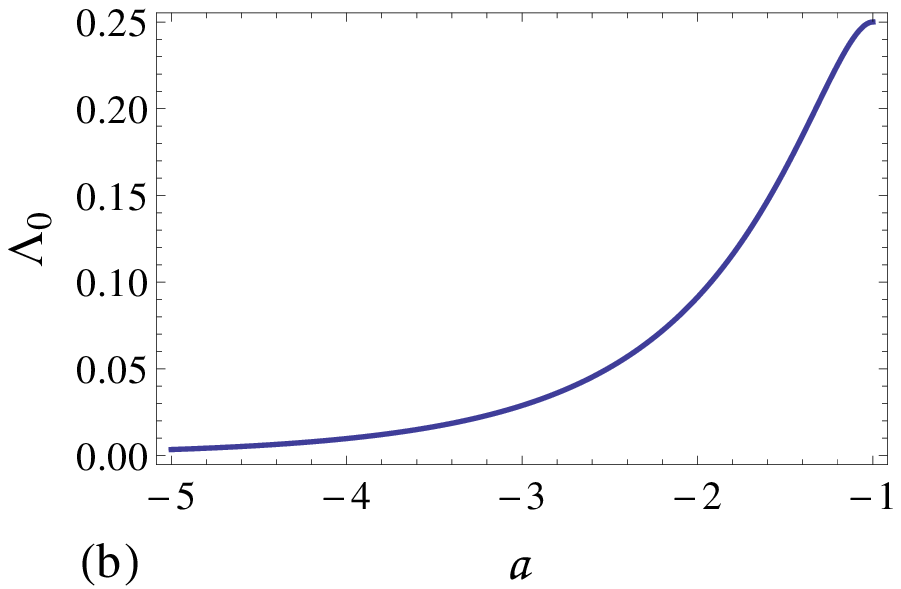}
\end{center}
\caption{(Color online) (a) Graphical solution of Eq.~(\ref{ccd}) in terms of the
convex function $ \exp(2a\lambda) $ and the straight line $ 1-2\lambda $. \yaming{As examples, $ a=-0.5 $ and $ a=-2 $ are used here to illustrate the shapes of the function $ \exp(2a\lambda) $ for the two phases $ a>-1 $ and $ a<-1 $, respectively.}
(b) The discrete eigenvalue $ \Lambda_0 $ for $ a<-1 $. When $ a=-1 $, the discrete eigenvalue merges with the continuous spectrum $ \Lambda\geq 1/4 $. }
\label{FigFunction}
\end{figure*}

For $ \Lambda<1/4 $, the sole eigenfunction is given by
[see Eqs.~(\ref{gc0}) and (\ref{gd})]
\begin{equation}\label{ccb}
u_{\Lambda}(v)=\left\{
\begin{array}{ll}
2\lambda e^{-(\lambda+1/2) v} &\quad \mbox{for } v>0\\
(2\lambda-1)e^{-(\lambda -1/2)v}&\\
\qquad +e^{(\lambda+1/2) v} & \quad \mbox{for } a<v<0,
\end{array}\right.
\end{equation}
where $ \lambda=\sqrt{1/4-\Lambda}>0 $.
The discrete eigenvalue is determined by the absorbing boundary condition (\ref{bf0}), which results in
\begin{equation}\label{ccd}
e^{2 a\lambda}=1-2\lambda \quad\mbox{for } \lambda>0.
\end{equation}
It is obvious that Eq.~(\ref{ccd}) has no real solution for $ \lambda $ in the region $ [1/2,\infty) $. Hence we have $ \Lambda>0 $ and can confine ourselves to search the solution of Eq.~(\ref{ccd}) for $ \lambda $ in the region $ (0,1/2) $.
Since $ \exp(2a\lambda) $ is convex as a function of $ \lambda $ and the right
hand side of Eq.~(\ref{ccd}) is a straight line, it is easy to verify [see Fig. \ref{FigFunction}(a)] that Eq.~(\ref{ccd}) has no real solution in $ (0,1/2) $ when $ a\geq -1 $ and admits a unique solution, denoted by $ \lambda_0 $, when $ a<-1 $.
The unique eigenvalue $ \Lambda_0=1/4-\lambda_0^2 $ can be obtained numerically from Eq.~(\ref{ccd}), being a monotonic function of the parameter $ a $ [see Fig. \ref{FigFunction}(b)]. As an aside
we remark that the solution of Eq.~(\ref{ccd}) can be expressed in terms of
the main branch of the Lambert W function \cite{CoGoHaJeKn_ACM96} by
$\lambda_0=1/2-W[a\exp(a)]/(2a)$. The other quantities which enter
the FPT distribution are easily computed. For the normalization factor,
Eqs.~(\ref{bh}) and (\ref{ccb}) yield
\begin{eqnarray}
Z_{\Lambda_0} &=& \int_a^{\infty}u_{\Lambda_0}^2(v)e^{|v|}\dd v\nonumber\\
&=& \left[-e^{2 a \lambda_0} +(1/2-\lambda_0)^2 e^{-2 a \lambda_0}\right]/\lambda_0\nonumber\\
&& - 4\lambda_0+2(1+a) .
\label{cce}
\end{eqnarray}
The integral of the eigenfunction which enters the distribution
[see Eqs.~(\ref{bb}) and (\ref{bg})] is evaluated as
\begin{equation}\label{ccf}
\int_{a}^{\infty}u_{\Lambda_0}(v)\dd v=2e^{(1/2-\lambda_0) a} -e^{(1/2 + \lambda_0)a}/(1/2 + \lambda_0).
\end{equation}

For $ \Lambda>1/4 $, the eigenfunction can be obtained explicitly as [see Eqs.~(\ref{gc0}) and (\ref{gd0})]
\begin{eqnarray}\label{ca}
u_{\Lambda}(v)=\left\{
\begin{array}{lll}
\sin(\kappa a)  \sin(\kappa v)e^{-v/2}&&\\
\qquad + \kappa\sin[\kappa (v-a)]e^{-v/2} & & \mbox{for } v>0\\
\kappa \sin[\kappa (v-a)]e^{v/2} & & \mbox{for } a<v<0,
\end{array}
\right.
\end{eqnarray}
where $ \kappa=\sqrt{\Lambda-1/4}>0 $. Moreover, the normalization factor
in Eq.~(\ref{bh0}) is given by [see Eq.~(\ref{gf})]
\begin{equation}\label{cc0}
Z_{\Lambda}=\pi[\kappa^2+\kappa \sin(2a\kappa)+\sin^2(a\kappa)]/2,
\end{equation}
and the integral over the eigenfunction which enters Eq.~(\ref{bg})
is evaluated as
\begin{equation}\label{cca}
\int_a^{\infty}u_{\Lambda}(v)\dd v=\kappa ^2e^{a/2}/(1/4+\kappa ^2).
\end{equation}

Thus, the spectrum consists of a continuous part $\Lambda>1/4$ and
an additional discrete lowest eigenvalue $\Lambda_0$ for $a<-1$
which merges with the continuous spectrum at $a=-1$
[see Fig.~\ref{FigFunction}(b)]. Hence we expect qualitative changes to appear
at such a critical value.

By using Eqs.~(\ref{bb}) and (\ref{bg}) we obtain the distribution of the FPT
as follows
\begin{widetext}
\begin{eqnarray}
f(T,v_0) &=&\chi_{\{a\leq -1\}}
 \Lambda_0 u_{\Lambda_0}(v_0)e^{|v_0|-\Lambda_0 T}\int_a^{\infty}u_{\Lambda_0}(v)\dd v/Z_{\Lambda_0}
 \nonumber\\
& & + \frac{2}{\pi}e^{|v_0|-T/4+a/2}\int_0^{\infty} \kappa ^2 u_\Lambda (v_0)e^{-\kappa ^2T}  /[\kappa^2+\kappa \sin(2a\kappa)+\sin^2(a\kappa)]\dd \kappa,
 \label{cb}
\end{eqnarray}
\end{widetext}
where
$ \chi_{\{a\leq -1\}} $ denotes the indicator function of the set $\{a\leq -1\}$, $ u_{\Lambda_0}(v_0) $
the eigenfunction of the discrete eigenvalue (\ref{ccb}), and $ u_{\Lambda}(v_0) $ the eigenfunction of the continuous part of the spectrum (\ref{ca}). The normalizations $ Z_{\Lambda_0} $ and $ \int_a^{\infty}u_{\Lambda_0}(v)\dd v $ are given in Eqs.~(\ref{cce}) and (\ref{ccf}), respectively.
In the trivial case $ a=0 $ the discontinuity does not enter the FPT problem
and the pure dry friction model is equivalent to that of the one-dimensional Brownian motion with constant drift \cite{KearneyMajumdar2005FirstPassage}. In such a case, the first term in Eq.~(\ref{cb}) does not contribute and the
integral can be evaluated in closed analytic form to yield
\begin{eqnarray}
\!\!\! f(T,v_0) &=& \frac{2}{\pi}e^{v_0/2-T/4}\int_0^{\infty}
\kappa \sin(\kappa v_0) e^{-\kappa ^2T} \dd \kappa
\nonumber\\
& = &
\frac{1}{2\sqrt{\pi}}\frac{v_0}{T^{3/2}}e^{-(v_0-T)^2/(4T)} \quad\mbox{for } v_0>0,\label{cf0}
\end{eqnarray}
a result which is consistent with Refs.~\cite{KearneyMajumdar2005FirstPassage,MajumdarComtet2002Sinai}.
Apart from this trivial case it seems to be difficult
to obtain a closed analytic expression from the representation (\ref{cb}).

\begin{figure*}
\begin{center}
\includegraphics[scale=0.9]{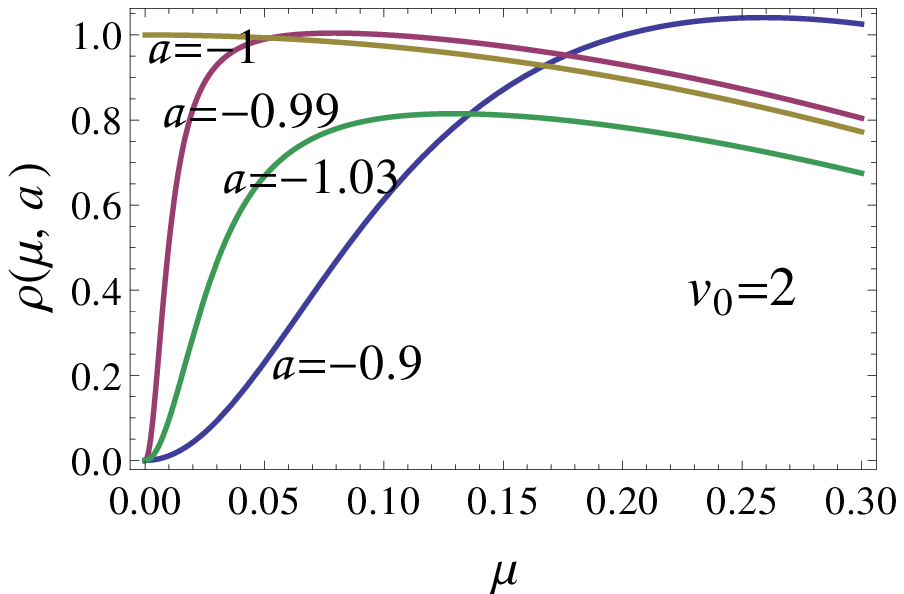}
\includegraphics[scale=0.9]{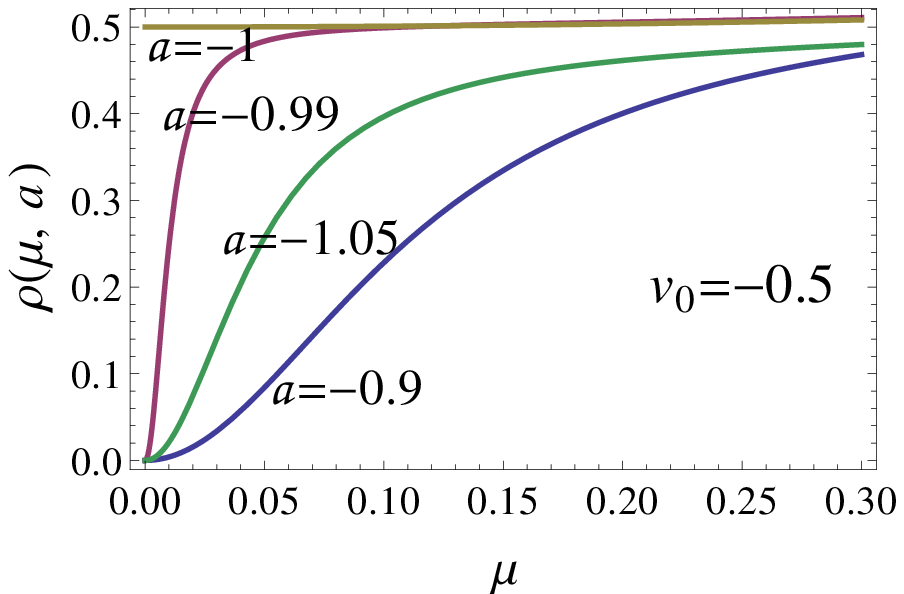}
\end{center}
\caption{(Color online) The kernel $ \rho(\kappa,a) $ [see Eq.~(\ref{aad})] appearing in the spectral decomposition (\ref{cb}) for
two different values of $v_0$ and various values of the exit point $a$. Here $ u_\Lambda (v_0) $ is given by
Eq.~(\ref{ca}). }\label{FigAsym}
\end{figure*}

Certainly the FPT distribution changes qualitatively at $a=-1$ when the
contribution in Eq.~(\ref{cb}) coming from the discrete eigenvalue comes
into play. That can be demonstrated by focussing on the tail behavior of
the distribution which in itself is of interest when rare events are of
interest. First of all it is obvious that for $a<-1$ the first term
in Eq.~(\ref{cb}) determines the decay which is plainly exponential
$ \exp(-\Lambda_0 T) $. For $a\geq -1$, the first term in Eq.~(\ref{cb}) vanishes, as
the coefficient of the characteristic function vanishes at $a=-1$, and the tail is determined
by evaluating the Laplace-type integral in the second term. If we have a closer look at the kernel
entering the Laplace-type integral
\begin{equation}\label{aad}
\rho(\kappa,a)=\kappa ^2 u_\Lambda (v_0) /[\kappa^2+\kappa \sin(2a\kappa)+\sin^2(a\kappa)],
\end{equation}
it is evident that for $a>-1$ the properties
\begin{eqnarray}
&& \lim_{\kappa\rightarrow 0}\rho(\kappa,a)=0,\\
&& \lim_{\kappa\rightarrow 0}\partial_{\kappa}\rho(\kappa,a)= 0,\\
&& \lim_{\kappa\rightarrow 0}\partial_{\kappa}^2\rho(\kappa,a)\neq 0
\end{eqnarray}
hold (see Fig.~\ref{FigAsym}). Hence it is straightforward to evaluate the Laplace-type integral to obtain a decay as
$ T^{-3/2}\exp(-T/4) $ for $ a>-1 $. For the critical case $ a=-1 $ the situation differs as
\begin{equation}\label{cb1}
    \lim_{\kappa\rightarrow 0}\rho(\kappa,-1)=\left\{
        \begin{array}{lll}
          1 & & \mbox{for } v_0>0\\
          1+v_0 & & \mbox{for } -1<v_0<0
        \end{array}
    \right.
 \end{equation}
holds. Here the Laplace method yields  $ T^{-1/2}\exp(-T/4) $ for $ a=-1 $.
To summarize, in the long time limit we have
\begin{equation}\label{cb2}
f(T,v_0)\sim\left\{
     \begin{array}{lll}
        e^{-\Lambda_0 T} & & \mbox{for } a<-1\\
        T^{-1/2}e^{-T/4} & & \mbox{for } a=-1\\
        T^{-3/2}e^{-T/4} & & \mbox{for } a>-1.
     \end{array}
\right.
\end{equation}

To obtain closed analytic expressions for the FPT distributions we
alternatively can resort to the Laplace transform of the backward
Kolmogorov equation. In this pure dry friction case
Eq.~(\ref{bq}) reads [see Eq.~(\ref{ca0})]
\begin{equation}\label{cf}
\frac{\partial^2 }{\partial v_0^2}\tilde{f}(s,v_0)-\sigma(v_0)\frac{\partial }{\partial v_0}\tilde{f}(s,v_0)-s\tilde{f}(s,v_0)=0,
\end{equation}
where the solution has to satisfy  the boundary conditions (\ref{br})
and (\ref{bra}) as well as
the matching condition (\ref{bs}) at $ v_0=0 $. It is in fact rather straightforward
to compute the solution of this linear second order problem and we
end up with
\begin{widetext}
\begin{equation}\label{cg}
 \tilde{f}(s,v_0)=\left\{
\begin{array}{lll}
\exp\big\{[\sqrt{1+4s}(a-v_0)+a+v_0]/2\big\}\sqrt{1+4s}/\theta(s,a) & & \mbox{for } v_0>0
\\
\exp[(1+\sqrt{1+4s})(a-v_0)/2] \theta(s,v_0)
/\theta(s,a) & & \mbox{for } a<v_0<0,
\end{array}
\right.
\end{equation}
\end{widetext}
where we have introduced the abbreviation
\begin{equation}
\theta(s,a)=\exp\left(a\sqrt{1+4s}\right)+\sqrt{1+4s}-1
\end{equation}
for the contribution appearing mainly in the denominator. Clearly
Eq.~(\ref{cg}) has a branch cut at $s=-1/4$ which relates with the
continuous spectrum found previously. In addition, the condition
$ \theta(s,a)=0 $, which is equivalent to Eq.~(\ref{ccd}),
determines a pole for $a<-1$.
Hence, when $ a<-1 $ the simple pole dominates the FPT distribution
in the tail to yield an exponential decay \cite{WhitehouseEvansMajumdar2013}.
Overall, the analytical structure
of the Laplace transform reflects the spectral properties mentioned
previously.

\begin{figure*}
\begin{center}
\includegraphics[scale=0.9]{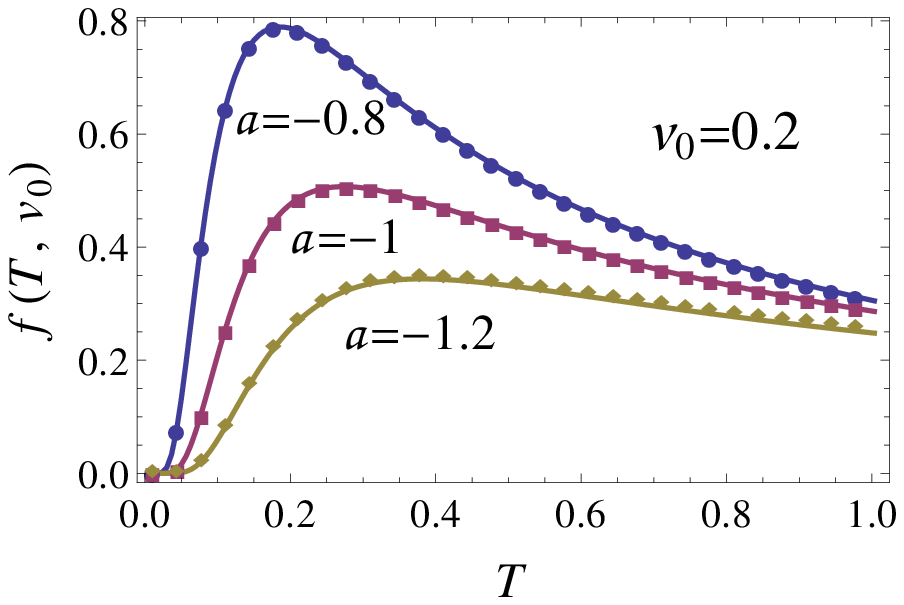}
\includegraphics[scale=0.9]{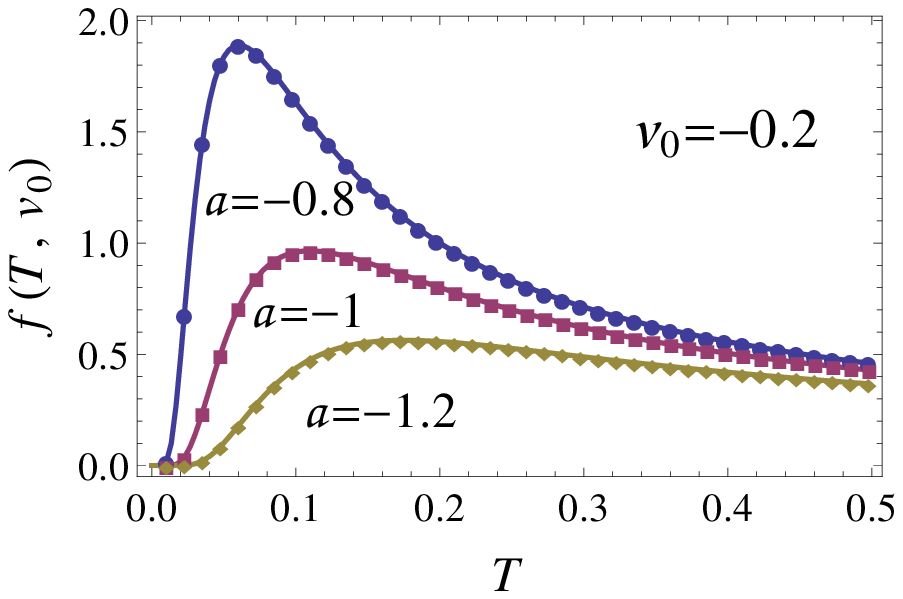}
\end{center}
\caption{(Color online) The FPT distribution of the pure dry friction case [see Eq.~(\ref{ca0})] for two values of initial velocity, \yaming{$ v_0=0.2 $} and $ v_0=-0.2 $, and different escape ranges. Lines correspond to a numerical inversion of Eq.~(\ref{cg}), and points to the evaluation of Eq.~(\ref{cb}).   }
\label{FigpureFPT}
\end{figure*}

\begin{figure}
\begin{center}
\includegraphics[scale=0.9]{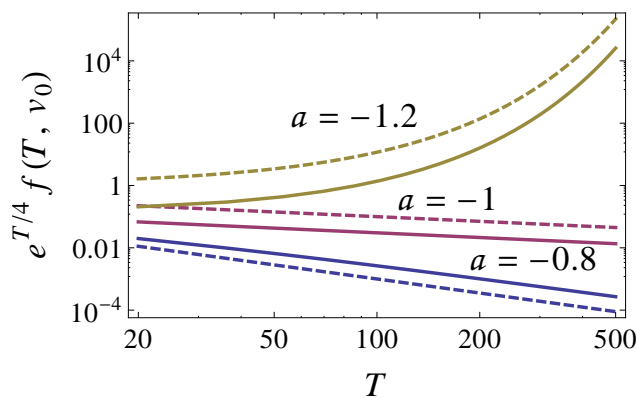}
\end{center}
\caption{(Color online) Comparison of the FPT distribution obtained from Eq.~(\ref{cb}) (solid) with the asymptotic result (\ref{cb2}) (dashed) for the initial velocity $ v_0=-0.2 $ and different escape ranges. Data are plotted on a doubly
logarithmic scale to uncover the power law corrections to the leading exponential behavior.}
\label{FiglogFPT}
\end{figure}

The inverse Laplace transform of Eq.~(\ref{cg}) does not seem
to be available in closed analytic form. As before, only the trivial case $a=0$
can be handled with ease which then results in Eq.~(\ref{cf0}). For the
other cases we resort to a so-called Talbot method \cite{Talbot1979Laplace,AbateValkoLaplace2004,AbateWhittLaplace2006}  to compute the FPT distribution in the time domain \footnote{A Mathematica implementation of this method is available at\\ http://libray.wolfram.com/infocenter/MathSource/5026/}. Fig.~\ref{FigpureFPT} shows that the expressions (\ref{cb}) and (\ref{cg}) give identical results, as expected. In addition, evaluation of those expressions confirm as well the asymptotic
decay given by Eq.~(\ref{cb2}) (see Fig.~\ref{FiglogFPT}).

\begin{figure}
\begin{center}
\includegraphics[scale=0.9]{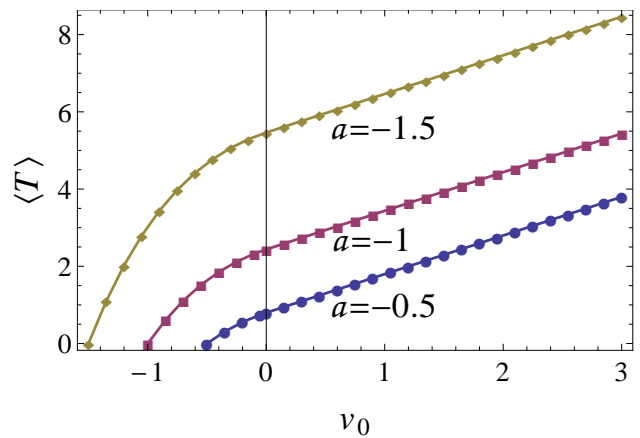}
\end{center}
\caption{(Color online) The MFPT $\langle T \rangle$ for different escape ranges.
Lines correspond to the analytic result (\ref{ch}), and points to a numerical evaluation of the first moment by using the spectral representation (\ref{cb}).
}\label{FigMFPT}
\end{figure}

The closed form of the characteristic function (\ref{cg}) allows us
to obtain easily the moments of the FPT via Eq.~(\ref{bt}). For the first moment,
i.e., for the mean first-passage time (MFPT) we have
\begin{equation}\label{ch}
\langle T \rangle=
\left\{
  \begin{array}{lll}
    2e^{-a}+a+v_0-2 & & \mbox{for }v_0>0\\
    2e^{-a}+a-v_0-2e^{-v_0} & & \mbox{for } a<v_0<0 .
  \end{array}
\right.
\end{equation}
The first moment clearly displays a transition when the initial condition
changes sign (see also Fig.~\ref{FigMFPT}). For $ v_0>0 $ the MFPT depends
linearly on the initial velocity. No particular feature is visible
at the transition at $a=-1$, as a change in the tail behavior has no
impact on the low order moments of the distribution.

\section{Biased Brownian motion with dry and viscous friction}
\label{sect4}
In this section, we consider the full model (\ref{aa}) and set $ \gamma=D=1 $ without loss of generality. Other cases can be covered by using the appropriate rescaling
\begin{equation}\label{da0}
x=\left( \gamma/D \right)^{1/2}v, \qquad \tau=\gamma t.
\end{equation}
Thus the model (\ref{aa}) can be written as Eq.~(\ref{ba}) with
\begin{equation}\label{da}
\Phi(v)=(|v|+\mu)^2/2-bv.
\end{equation}
The corresponding eigenvalue problem (\ref{bf}) with potential (\ref{da}) can be solved by using parabolic cylinder
functions \cite{Buchholz1969Hypergeometric}, which are denoted by $ D_{\nu}(z) $. For the convenience of the reader
we summarize the details of the derivation
in Appendix \ref{appB}.

The eigenvalues are discrete and determined by the characteristic equation [see Eq.~(\ref{ij})]
\begin{eqnarray}
&&\!\!\!\!\!\!\!\!\!\!\!\!\!\!\!\! \bar{\theta}(\Lambda,a,\mu,b)
\nonumber\\
&=&\Gamma(1-\Lambda)
\{
[D_{\Lambda}(\mu+b)D_{\Lambda-1}(\mu-b)
\nonumber\\
&&+D_{\Lambda}(\mu-b)D_{\Lambda-1}(\mu+b)]
 D_\Lambda(a-\mu -b) \nonumber\\
 &&
-
[D_{\Lambda}(-\mu-b)D_{\Lambda-1}(\mu-b)
\nonumber\\
 && -D_{\Lambda}(\mu-b)D_{\Lambda-1}(-\mu-b)]D_\Lambda(-a+\mu +b)
\}
\nonumber\\
 &=& 0.
\label{de}
\end{eqnarray}

For $ \mu=0 $, the model considered here reduces to the Ornstein-Uhlenbeck process and the characteristic
equation (\ref{de}) simply reads
\begin{equation}\label{de1}
D_\Lambda(a-b)= 0,
\end{equation}
which agrees with the standard result of the Ornstein-Uhlenbeck process (see for instance Ref.~\cite{Siegert1951FPT}).
It is furthermore unexpected that the characteristic equation (\ref{de1}) coincides with those of the odd part of the spectrum for a Fokker-Planck equation subjected to dry and viscous friction only
\cite{TouchetteStraeten2010Brownian}.
While odd eigenfunctions vanish at the origin and thus fulfil some kind of absorbing boundary condition it is
not intuitively obvious why the argument of the parabolic cylinder function is in one case the absorbing
boundary and in the other case the dry friction itself.

\begin{figure*}
\begin{center}
\includegraphics[scale=0.9]{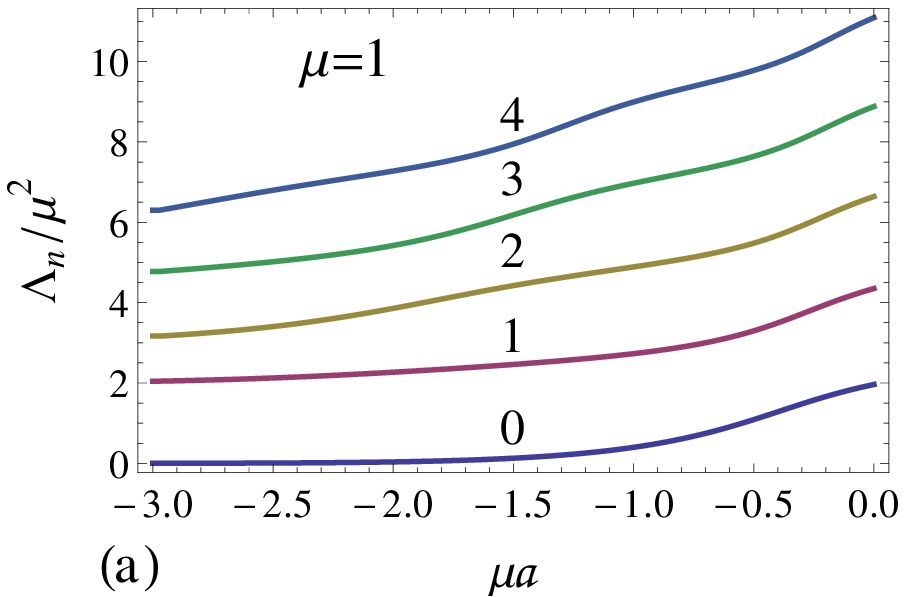}
\includegraphics[scale=0.9]{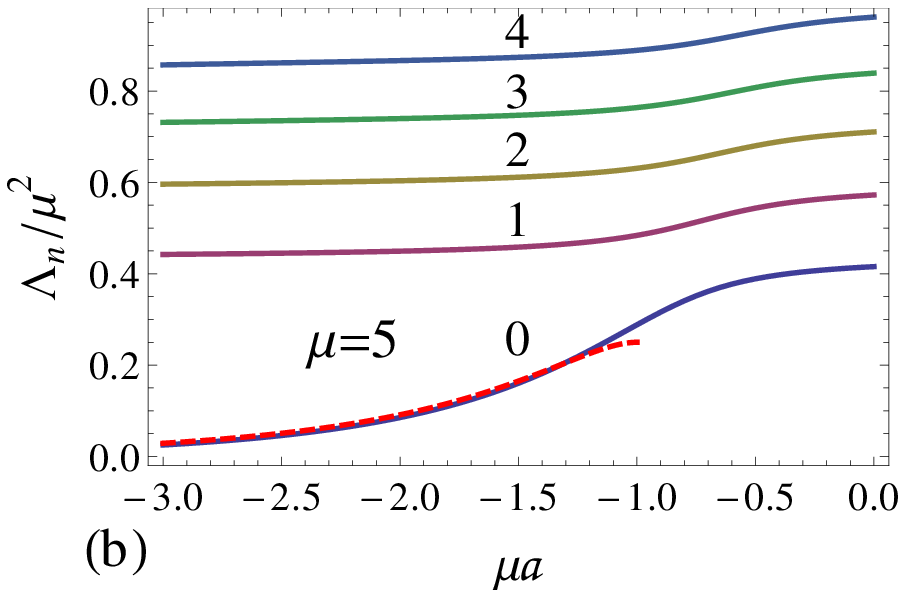}
\end{center}
\caption{(Color online) The first five rescaled eigenvalues $\Lambda_n/\mu^2$ for the model without bias ($b=0$) as a function of the rescaled exit point
$\mu a$ for two different values of $ \mu $,
according to Eq.~(\ref{de}). The dashed line in (b) depicts the discrete branch of the
model with dry friction only [see Fig.~\ref{FigFunction}(b)].}
\label{FigViscousDryDe}
\end{figure*}

To link the current result with the previous section let us first consider the special case without bias ($ b=0 $). Intuitively, we expect that if the dry friction term dominates the viscous friction force then
the particle will behave like the one subjected to dry friction only. Hence the spectrum obtained from
Eq.~(\ref{de}) for large values of $\mu$  should resemble the spectrum described in the previous section
[see, e.g., Fig.~\ref{FigFunction}(b)]. In particular it means that a large gap should develop between the lowest eigenvalue
and a quasicontinuous part for small negative values of $ a $.
For comparison of the models with and without viscous friction
[see Eqs.~(\ref{ca0}) and (\ref{da})] we observe that a rescaling of the
velocity by $\mu$ and of time by $\mu^2$
transforms the stochastic differential equation with dry and viscous friction
to the model with dry friction and a small viscous part of $O(1/\mu^2)$ which vanishes in the limit
$\mu\rightarrow\infty$. Thus, to compare the eigenvalues obtained
from the characteristic equation (\ref{de}) with the spectrum computed in the
previous section we rescale velocities
by $\mu$ and eigenvalues by $1/\mu^2$. Then,
indeed
numerical evaluation of Eq.~(\ref{de}) confirms what one expects intuitively (see Fig.~\ref{FigViscousDryDe}).
The eigenvalues as a function of the exit position $a$ develop a gap if $\mu$ is sufficiently large,
even though the transition is smoothened by the finite viscous friction. If dry and viscous friction
become comparable, i.e., if $\mu$ becomes too small such a feature is going to disappear.

\begin{figure}
\begin{center}
\includegraphics[scale=0.9]{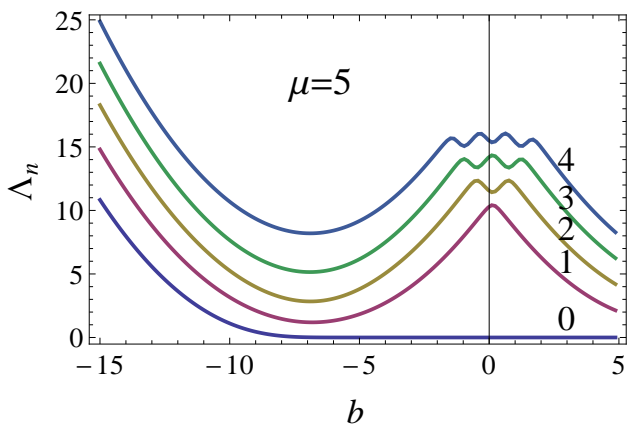}
\end{center}
\caption{(Color online) The first five eigenvalues as a function of the bias $b$ for
exit point at $a=-5$ and  dry friction coefficient $\mu=5$,
obtained from Eq.~(\ref{de}). The stick-slip transition, i.e., the
narrowing of the spectral gap at $b=\pm\mu=\pm 5$ and the transition
to a ballistic exit at $b=-\mu+a=-10$ are clearly visible.}
\label{FigFullSpectral}
\end{figure}

If we impose a force on the particle the finite bias will cause a stick-slip
transition at $|b|=\mu$ where the minimum of the potential
(\ref{da}), i.e., the
deterministic stationary state, changes from vanishing to finite velocity.
The characteristics of such a transition are reflected by the eigenvalue
spectrum as well (see Fig.~\ref{FigFullSpectral}). For small value of the bias,
$|b|<\mu$, a case which we will call for brevity the dry phase,
a substantial spectral gap appears between the lowest and
the subleading eigenvalues. This gap shrinks when the transition
at $|b|=\mu$ is approached.
The spectral gap corresponds to a fast decay of velocity correlations in the system
with small bias (see Ref.~\cite{TouchetteStraeten2010Brownian}).
If the bias is sufficiently negative, i.e., $b<-\mu$, a case which we will call the wet phase, the potential (\ref{da}) develops a quadratic minimum
and the spectrum resembles that of the Ornstein-Uhlenbeck process.
As with regards to the exit time problem a second transition will occur when
on decreasing the force further the quadratic minimum of the
potential moves beyond the exit point at $b=-\mu+a$. Then the exit \yaming{from} the
region occurs in a purely ballistic way which decreases the exit time
considerably. Hence that transition is related with an
increase of the lowest eigenvalue (see Fig.~\ref{FigFullSpectral}).
These two transitions are clearly visible if the diffusion is sufficiently
small, i.e., $\mu$ sufficiently large. But they become obscured by noise for
large diffusion, i.e., if $\mu$ becomes too small. Finally, in the dry phase
the spectrum shows avoided level crossings for small bias,
which are reminiscent of spectral properties in nonintegrable dynamical systems.

\begin{figure}
\begin{center}
\includegraphics[scale=0.9]{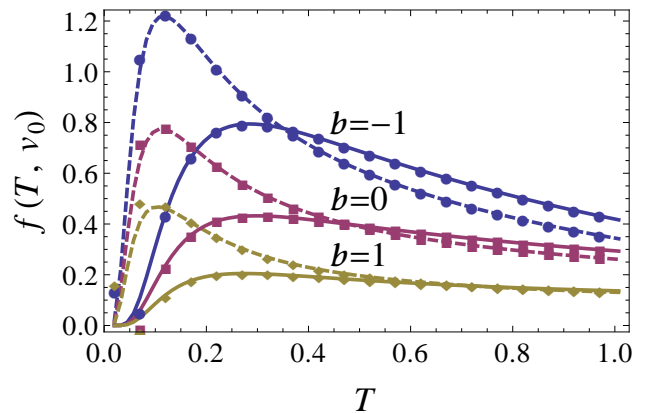}
\end{center}
\caption{(Color online) The distribution of the FPT for $\mu=1$, $a=-1$, two
values of initial velocity, $v_0=0.2$ (solid) and $v_0=-0.2$
(dashed), and different values of the bias $b$.
Lines correspond to a numerical inversion
of the Laplace transform (\ref{dg}), and points to the evaluation
of Eq.~(\ref{df}) taking the first twenty modes into account. A larger number of modes would
be required to reproduce the exact result for very small values of $t$.}
\label{FigDFPTfull}
\end{figure}

As we have access to the entire spectrum we can derive from Eqs.~(\ref{bb}) and (\ref{bg}) the FPT distribution
\begin{equation}\label{df}
f(T,v_0)=e^{\Phi(v_0)}\sum_{\Lambda} \Lambda u_{\Lambda}(v_0)e^{-\Lambda T }\int_a^{\infty} u_{\Lambda}(v)\dd v/Z_{\Lambda},
\end{equation}
where the sum is taken over all the discrete eigenvalues [see Eq.~(\ref{de})], $u_{\Lambda}(v_0)$
refers to the eigenfunction given by Eq.~(\ref{ib0}), the integral  $\int_a^{\infty} u_{\Lambda}(v) \dd v$
is stated in Eq.~(\ref{il}) and the normalization factor $Z_\Lambda$ is given by Eq.~(\ref{in}).
It is thus straightforward to evaluate the shape of the distribution function (see, e.g., Fig.~\ref{FigDFPTfull}).
While it seems to be difficult to obtain a closed analytic expression for this distribution we may pursue the
approach used in the previous section and focus on the Laplace transform. In fact, Eq.~(\ref{bq}) tells us that
[see Eq.~(\ref{da})]
\begin{equation}\label{dg0}
\frac{\partial^2 }{\partial v_0^2}\tilde{f}(s,v_0)-(v_0+\mu\sigma(v_0)-b)\frac{\partial }{\partial v_0}\tilde{f}(s,v_0)-s\tilde{f}(s,v_0)=0,
\end{equation}
where the Laplace transform has to obey the boundary conditions (\ref{br})
and (\ref{bra}) as well as the matching condition (\ref{bs}).
Solving Eq.~(\ref{dg0}) is rather straightforward, as the boundary value
problem for the Laplace transform is the formally adjoint of the
eigenvalue problem [see Eqs.~(\ref{ia}) and (\ref{ib})]. It is well
known and easy to confirm that the solution
of the adjoint problem can be written in terms of the analytic expression
for the eigenfunction (see Ref.~\cite{Gardiner1990StoMeth}) if we multiply the eigenfunction with an exponential factor
$\exp[\Phi(v_0)]$ containing the potential (\ref{da}).
Thus, the solution of Eq.~(\ref{dg0}) can be written down directly as
\begin{equation}\label{dg}
\tilde{f}(s,v_0)=\frac{e^{(a-\mu-b)^2/4-\Phi(a)}}{\bar{\theta}(-s,a,\mu,b)}u_{-s}(v_0)e^{\Phi(v_0)}\quad \mbox{for } v_0>a,
\end{equation}
where $ u_{-s}(v_0) $ refers to Eq.~(\ref{ib0}), and the additional
normalization factor containing the characteristic equation (\ref{de})
is obtained by using the boundary condition (\ref{br}).
Obviously the poles of the Laplace transform are determined by
the characteristic equation (\ref{de}) and thus
reflect the spectral structure discussed previously. In addition,
the smallest simple pole determines the exponential tail of $ f(T,v_0) $.

As stated before, for $ \mu=0$ the model investigated here corresponds to
the exit time problem of the Ornstein-Uhlenbeck process, which has been paid
much attention to in the past (see for instance
Refs.~\cite{AliliPatiePedersen2005ou,WangUhlenbeck1945Brownian,Siegert1951FPT,DarlingSiegert1953fpt,BlakeLindsey1973ou,Leblanc1998ou}). In this case
Eq.~(\ref{dg}) simplifies considerably
and reads [see Eqs.~(\ref{de}) and (\ref{ib0})]
\begin{equation}
\tilde{f}(s,v_0)=\frac{e^{(v_0-b)^2/4}D_{-s}(v_0-b)}{e^{(a-b)^2/4}D_{-s}(a-b)} \quad \mbox{for } v_0>a,
\end{equation}
which is consistent with the standard result stated, for instance,
in Ref.~\cite{Siegert1951FPT}.

The analytic expressions Eqs.~(\ref{df}) or (\ref{dg}) now allow us to
discuss the dependence of the exit time problem on the initial velocity $v_0$.
Both expressions, if properly evaluated, give of course identical
results (see Fig.~\ref{FigDFPTfull}). Here we are going to pay
particular attention to the impact of the discontinuity
appearing at the origin. Depending on the sign of the initial velocity
the particle has to pass the discontinuity at $v=0$ before exiting at $a<0$.
Thus, a qualitative change of the FPT distribution is expected depending
on the sign of $v_0$. In fact, such a feature is already visible from
Eq.~(\ref{dg}), as different analytical branches of the eigenfunction
(\ref{ib0}) come into play if $v_0$ changes sign. The dependence on $v_0$
is still smooth but not differentiable of higher order.
The FPT distributions for small  positive and small negative values of
$v_0$ look distinctively different,
as shown in Fig.~\ref{FigDFPTfull}. For $v_0>0$ the particle
has to pass through $v=0$ before exiting and thus sticks at the origin
at least if the bias is small, causing larger exit times. Thus, the
distribution overall is shifted to the right, compared to the case $v_0<0$.

\begin{figure*}
\begin{center}
\includegraphics[scale=0.9]{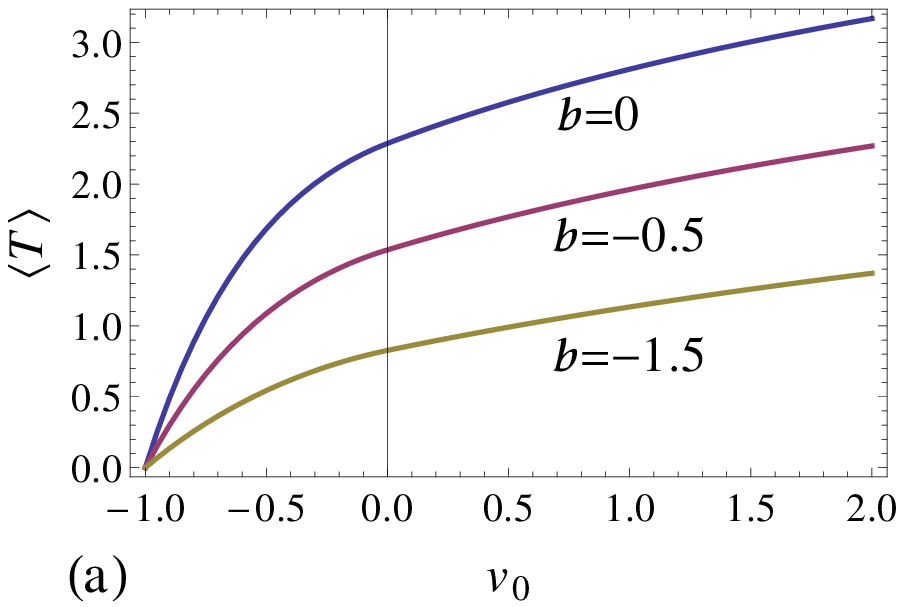}
\includegraphics[scale=0.885]{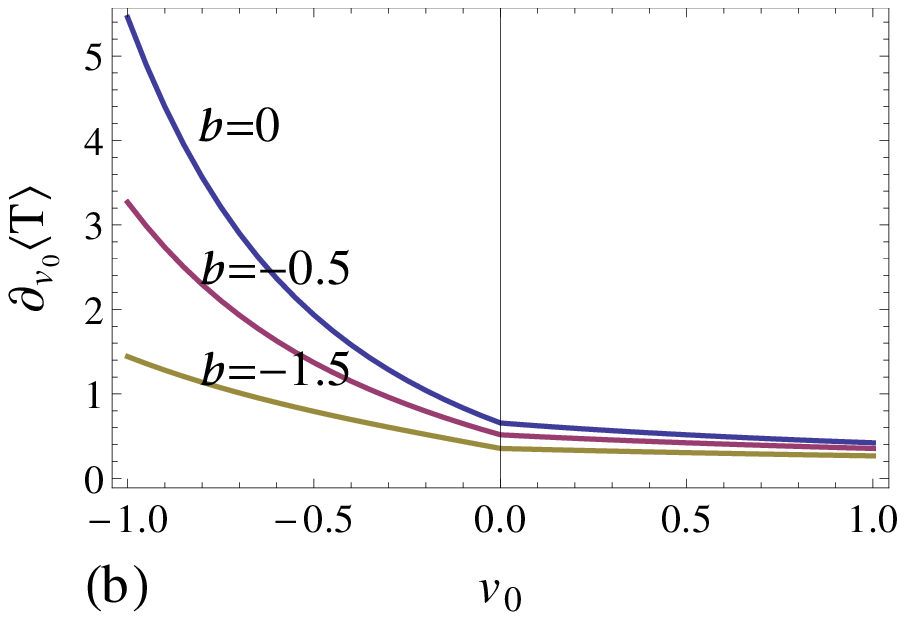}
\end{center}
\caption{(Color online) (a) MFPT $\langle T\rangle$ as a function of the initial value $v_0$ for
$\mu=1$, exit condition $a=-1$, and different values of the bias,
covering the dry phase  $|b|<\mu$ as well as the wet phase $b<-\mu$.
(b)  First derivative of the MFPT with respect to the initial value for the
same data.}\label{FigFullMFPT}
\end{figure*}

The just mentioned phenomenon can be better illustrated by looking
at the MFPT which can be obtained in closed analytic form via Eqs.~(\ref{bt}) and
(\ref{dg}) even for very small values of the diffusion, i.e., for large
values of $\mu$. While the analytic expression can be written down
we just refer to the graphical evaluation of the expressions (see Fig.~\ref{FigFullMFPT}).
For small bias, $ |b|<\mu $, i.e., in the dry phase there is
a possibility that the particle sticks at the origin which will impact on the
MFPT. If the particle starts at $v_0<0$ it has less chance to stick at the origin when $v_0$ becomes smaller,
and the change of the MFPT with regards to $v_0$ becomes fairly large.
On the contrary, if we choose a positive initial
velocity $ v_0>0 $, the particle has always to pass $ v=0 $ before exiting at $ a<0 $. Thus no huge
variation of the MFPT with $v_0$ is detected. If we decrease the bias and enter the wet phase $ b<-\mu $,
the particle does not stick any more and the just mentioned feature almost disappears.
This scenario is much more pronounced if we look at the first derivative $ \partial_{v_0}\langle T \rangle $ [see Fig.~\ref{FigFullMFPT}(b)]. Like the distribution function itself the MFPT is
continuously differentiable, but loses analyticity due to the discontinuity at $ v_0=0 $.
A kink can be seen clearly at the origin for small bias $ |b|<\mu $, which
separates the two different regimes of the MFPT for negative and positive initial velocities.
This feature is suppressed if
we decrease the bias and finally enter the wet phase with $ b<-\mu $ where the kink
almost disappears.

\section{Conclusion}
\label{sect5}

In this paper we have studied the FPT problem of Brownian motion with dry and
viscous friction. There has been renewed interest in such exit time problems from two different
points of view. On the one hand prediction and forecasting of extreme events and the related large deviation theory are closely related to exit time problems.
On the other hand, the particular setup studied here is a special example of a piecewise-smooth dynamical system. While such systems are extensively used in engineering sciences only
recently the attempt has been made to put this subject in the systematic framework of dynamical system's theory.

As a case study we have considered here
a simple piecewise-linear model which can be largely solved by
analytical means.
In physicists terms we have considered a particle
subjected to dry and viscous friction, to noise, and to an external force.
This is one of the few models for which the FPT distribution can be obtained analytically either by solving the Fokker-Planck equation
via a spectral decomposition method or by solving the backward Kolmogorov equation in the Laplace space.
While the first method gives more insight into the underlying dynamical mechanisms through the additional spectral information, the second is able to deliver closed analytic expressions
for the MFPT.

The simplest case, where only dry friction acts on the particle, already shows one of the main features, a phase
transition phenomenon in the spectrum which is related to the position of the exit point. A unique discrete eigenvalue
links up with the continuous part of the spectrum at a critical size of the exit region. Such a transition translates into
different asymptotic properties of the FPT distribution.
The signature of this transition persists if the viscous friction and the external bias are taken into account,
even though the transition is blurred by the finite diffusion. In this full model two new features
occur, i.e., a stick-slip transition and a transition to a ballistic exit of the particle. All three transitions
are clearly visible in the discrete spectrum of the full model, especially at low diffusion, signalling
the different rates of asymptotic decay of the FPT distribution. As an aside, the analysis of this model covers
as special cases the Ornstein-Uhlenbeck process on the one hand, and the previously discussed dry friction
case on the other.

The availability of analytical results for higher dimensional stochastic models is rather limited,
contrary to the one-variable case. Even the computation of the stationary distribution is often a challenge
if detailed balance is violated, and dynamical quantities, like correlations or
exit probabilities are certainly out of reach. Having said that, models with more than one degree of freedom are
prevalent in applications and any progress
on the analytical side is certainly welcomed, even if simple model systems are considered. In that sense
the inclusion of inertia in the model discussed here is a rewarding goal, which could lead to
predictions that are experimentally relevant and could trigger corresponding experimental
investigations. Progress in that direction seems
possible even though the analysis may not be entirely straightforward.

\begin{acknowledgments}
Y.C. was \yaming{supported} by the China Scholarship Council and NUDT's Innovation Foundation (Grant No. B110205). W.J. gratefully acknowledges support from EPSRC through Grant No. EP/H04812X/1 and DFG through SFB910. We would also like to thank Hugo Touchette for the useful discussions on large deviation theory of Brownian motion.
\end{acknowledgments}

\appendix

\section{Eigenvalue problem for the inviscid case}
\label{appA}

Without viscous damping and driving  Eq.~(\ref{bf}) reads
[see Eq.~(\ref{ca0})]
\begin{eqnarray}
-\Lambda u_\Lambda(v)=u_\Lambda'(v)+u_\Lambda''(v) && \quad \mbox{for } v>0 \label{ga}\\
-\Lambda u_\Lambda(v)=-u_\Lambda'(v)+u_\Lambda''(v) && \quad \mbox{for } a<v<0. \label{gb}
\end{eqnarray}
Let
\begin{equation}\label{gc0}
u_\Lambda(v)=e^{-|v|/2}\varphi_\Lambda(v),
\end{equation}
then Eqs. (\ref{ga}) and (\ref{gb}) can be written as
\begin{equation}\label{gc}
\varphi''_{\Lambda}(v)=(1/4-\Lambda)\varphi_{\Lambda}(v) \quad \mbox{for } v\neq 0.
\end{equation}

On the one hand, for $ \Lambda<1/4 $ let us introduce the positive variable $ \lambda=\sqrt{1/4-\Lambda} $. Then the solution of Eq.~(\ref{gc}) which results in a finite normalization factor [see Eq.~(\ref{bh})] is given by
\begin{equation}\label{gd}
\varphi_{\Lambda}(v)=\left\{
    \begin{array}{lll}
        A_\lambda e^{-\lambda v} & & \mbox{for } v>0
          \\
        B_\lambda e^{\lambda v}+C_\lambda e^{-\lambda v} & & \mbox{for } a<v<0.
    \end{array}
\right.
\end{equation}
Choose $ A_\lambda=2\lambda $ and use the matching conditions (\ref{bi}) and (\ref{bj}) to determine the other two coefficients in Eq.~(\ref{gd}) as
\begin{eqnarray}
B_\lambda=1, \qquad C_\lambda=2\lambda-1.
\end{eqnarray}
The eigenvalue is now determined by the absorbing boundary condition
(\ref{bf0}), i.e.,  $ \varphi_{\Lambda}(a)=0 $, which results in Eq.~(\ref{ccd}).

On the other hand, for $ \Lambda>1/4 $ the solution of Eq.~(\ref{gc})
which vanishes at $v=a$, i.e., which satisfies the absorbing boundary
condition (\ref{bf0}), is given by
\begin{equation}\label{gd0}
\varphi_\Lambda(v)=\left\{
    \begin{array}{lll}
       \bar{A}_\kappa\sin(\kappa v)+\bar{B}_\kappa\cos(\kappa v) & & \mbox{for } v>0
         \\
       \bar{C}_\kappa \kappa\sin[\kappa(v-a)] & & \mbox{for } a<v<0,
    \end{array}
\right.
\end{equation}
where we have introduced the abbreviation $ \kappa=\sqrt{\Lambda-1/4}>0 $.
Choose $ \bar{C}_\kappa=\kappa $, then by using the matching conditions
(\ref{bi}) and (\ref{bj}), the two parameters
$ \bar{A}_\kappa $ and $ \bar{B}_{\kappa} $ are evaluated as
\begin{eqnarray}
\bar{A}_\kappa  = \kappa  \cos(a\kappa )+\sin(a\kappa ), \qquad \bar{B}_\kappa  = -\kappa  \sin(a\kappa ).  \label{cc}
\end{eqnarray}
Hence Eq.~(\ref{ca}) follows from substituting Eq.~(\ref{gd0})
into Eq.~(\ref{gc0}).
For the normalization, Eqs.~(\ref{gc0}) and (\ref{gd0}) result in
\begin{widetext}
\begin{eqnarray}
&& \!\!\!\!\!\!\!\!\!\!\!\!\!\!\!\!\!\!\!\!\!  \int_{a}^{\infty} u_\Lambda(v)u_{\Lambda'}(v)e^{|v|}\dd v
\nonumber\\
 &=&\int_{0}^{\infty} [\bar{A}_\kappa\sin(\kappa v)+\bar{B}_\kappa \cos(\kappa v)][\bar{A}_{\kappa '}\sin(\kappa 'v)+\bar{B}_{\kappa '}\cos(\kappa 'v)]\dd v
+\int_{a}^{0} \kappa \kappa '\sin[\kappa (v-a)]\sin[\kappa '(v-a)]\dd v
\nonumber\\
 &=& \int_{0}^{\infty} \bigg\{\frac{1}{2}\big(\bar{A}_\kappa \bar{A}_{\kappa '}+\bar{B}_\kappa \bar{B}_{\kappa '}\big)\cos[(\kappa -\kappa ')v]
+ \frac{1}{2}\big(\bar{B}_\kappa\bar{B}_{\kappa '}-\bar{A}_\kappa\bar{A}_{\kappa '}\big)\cos[(\kappa +\kappa ')v]
\nonumber\\
 &&+ \frac{1}{2}\big(\bar{A}_\kappa\bar{B}_{\kappa '}-\bar{A}_{\kappa '}\bar{B} _{\kappa }\big)\sin[(\kappa -\kappa ')v]
 + \frac{1}{2}\big(\bar{A}_\kappa\bar{B}_{\kappa '}+\bar{A}_{\kappa '}\bar{B}_{\kappa }\big)\sin[(\kappa +\kappa ')v]\bigg\}\dd v
 -\frac{\kappa\bar{A}_\kappa\bar{B}_{\kappa '}-\kappa '\bar{A}_{\kappa '}\bar{B}_\kappa }{\kappa ^2-{\kappa '}^2}
\nonumber \\
 &=& \frac{\pi}{2}(\bar{A}_\kappa ^2+\bar{B}_\kappa ^2)\delta(\kappa -\kappa'),
\label{gf}
\end{eqnarray}
\end{widetext}
which shows that the normalization factor $ Z_{\Lambda} $ satisfies
Eq.~(\ref{cc0}) if we take Eq.~(\ref{cc}) into account.
To derive Eq.~(\ref{gf}), we have used the
standard identities for the $\delta$-- and the principal value distribution
\begin{equation}\label{gg}
\int_{0}^{\infty}\cos(\kappa v)\dd v=\pi \delta(\kappa), \qquad
\int_{0}^{\infty}\sin(\kappa v)\dd v=P\left(\frac{1}{\kappa}\right) .
\end{equation}

\section{Eigenvalue problem for the general case}
\label{appB}

For the model (\ref{aa}), the eigenvalue problem (\ref{bf}) reads
\begin{widetext}
\begin{eqnarray}
 -\Lambda u_\Lambda(v)=[ (v+\mu-b)u_\Lambda(v) ]'+u_\Lambda''(v) & & \mbox{ for } v>0 \label{ia}\\
 -\Lambda u_\Lambda(v)=[ (v-\mu-b)u_\Lambda(v) ]'+u_\Lambda''(v) & & \mbox{ for } a<v<0, \label{ib}
\end{eqnarray}
if we adopt the notation used for Eq.~(\ref{da}).
These two equations are a special case of the so-called Kummer's equation, which can be solved in terms of
parabolic cylinder functions \cite{TouchetteStraeten2010Brownian}.
The solution of Eqs.~(\ref{ia}) and (\ref{ib}) which vanishes at infinity is
given by (see Refs.~\cite{TouchetteThomas2012Brownian,Buchholz1969Hypergeometric})
\begin{equation}\label{ib0}
 u_\Lambda(v) =\left\{
   \begin{array}{lll}
    A_{\Lambda}e^{-(v+\mu-b)^2/4}D_\Lambda(v+\mu-b) & & \mbox{for } v>0
\\
    B_{\Lambda}e^{-(v-\mu-b)^2/4}D_\Lambda(v-\mu-b)
+C_{\Lambda} e^{-(v-\mu-b)^2/4}D_\Lambda(-v+\mu+b) & & \mbox{for } a<v<0,
\end{array}
\right.
\end{equation}
where $ D_{\Lambda} $ denotes the parabolic cylinder function. Here we have used a fundamental
system in terms of $D_{\nu} (z)$ and $D_{\nu}(-z)$ to write down the solution. Such a
fundamental system degenerates for $\nu$ being an integer. Thus, our expressions may
contain spurious singularities at integer values of $\Lambda$ which have to be taken care of.
The coefficients $ A_{\Lambda} $, $ B_{\Lambda} $ and $ C_{\Lambda} $
depend on the parameters $b$ and $\mu$ as well, but are independent of $ v $.

Using Eq.~(\ref{ib0}) the matching conditions (\ref{bi}) and (\ref{bj})
result in a set of linear homogeneous equations
\begin{eqnarray}
&& B_{\Lambda}D_{\Lambda}(-\mu-b)+C_{\Lambda}D_{\Lambda}(\mu+b)= e^{\mu b}A_{\Lambda}D_{\Lambda}(\mu-b),\label{ic1}\\
&& B_{\Lambda} D_{1+\Lambda}(-\mu-b)-
  C_{\Lambda}D_{1+\Lambda}(\mu+b)= e^{\mu b}A_{\Lambda}[D_{1+\Lambda}(\mu-b)-2\mu D_{\Lambda}(\mu-b)] \label{ic2}
\end{eqnarray}
\end{widetext}
when the property
\begin{equation}\label{ig}
\frac{\dd  e^{-z^2/4}D_{\nu}(z)}{\dd z}=-e^{-z^2/4}D_{\nu+1}(z)
\end{equation}
of the parabolic cylinder function is employed. For $ A_\Lambda $ we choose
\begin{equation}\label{ic0}
A_{\Lambda}=\sqrt{2\pi} e^{-\mu b}.
\end{equation}
Then, the other two coefficients in Eq.~(\ref{ib0}) follow as
\begin{eqnarray}
 B_{\Lambda}&=&-\Lambda\Gamma(-\Lambda) [D_{\Lambda}(\mu+b)D_{\Lambda-1}(\mu-b)
 \nonumber\\
&& +D_{\Lambda}(\mu-b)D_{\Lambda-1}(\mu+b)], \label{ic}\\
 C_{\Lambda}&=&\Lambda\Gamma(-\Lambda)[D_{\Lambda}(-\mu-b)D_{\Lambda-1}(\mu-b)
 \nonumber\\
 &&-D_{\Lambda}(\mu-b)D_{\Lambda-1}(-\mu-b)], \label{id}
\end{eqnarray}
where we have used the identities
\begin{eqnarray}
D_{\nu}(z)D_{\nu+1}(-z)+D_{\nu}(-z)D_{\nu+1}(z)=\frac{\sqrt{2\pi}}{\Gamma(-\nu)}, \label{ii}
\end{eqnarray}
\begin{equation}
zD_{\nu}(z)-D_{\nu+1}(z)-\nu D_{\nu-1}(z)=0 \label{ii1}
\end{equation}
to simplify the above two expressions.

The characteristic equation simply follows from the boundary condition (\ref{bf0}), and
is thus given by
\begin{equation}\label{ij}
B_{\Lambda}D_{\Lambda}(a-\mu-b)+C_{\Lambda}D_{\Lambda}(-a+\mu+b)=0.
\end{equation}
Using the identities (\ref{ii}) and (\ref{ii1}) we arrive at Eq.~(\ref{de}).

For the integral over the eigenfunction which enters the FPT distribution (\ref{df})
we obtain by using, e.g., the differential identity (\ref{ig})
\begin{widetext}
\begin{eqnarray}
 \int_a^{\infty} u_{\Lambda}(v) \dd v
&=&A_{\Lambda} e^{-(\mu-b)^2/4}D_{\Lambda-1}(\mu-b)
- e^{-(\mu+b)^2/4}[B_{\Lambda}D_{\Lambda-1}(-\mu-b)-C_{\Lambda}D_{\Lambda-1}(\mu+b)]
\nonumber\\
 &&
+ e^{-(a-\mu-b)^2/4}[B_{\Lambda}D_{\Lambda-1}(a-\mu-b)-C_{\Lambda}D_{\Lambda-1}(-a+\mu+b)].
 \label{il}
\end{eqnarray}

Finally to compute the  normalization let us consider the integral
\begin{eqnarray}
&&\!\!\!\!\!\!\!\!\!\!\!\!\!\!\!\!\!\!\!\!\! (\Lambda-\Lambda')\int_a^{\infty}  e^{(v+\mu\sigma(v))^2/2-bv}u_\Lambda(v)u_{\Lambda'}(v) \dd v
\nonumber\\
 &=&  e^{-\mu b-b^2/2}(\Lambda-\Lambda')\int_a^{0} [B_\Lambda D_\Lambda(v-\mu -b)
+C_\Lambda D_\Lambda(-v+\mu +b)]
\nonumber\\
&&\times [B_{\Lambda'}D_{\Lambda'}(v-\mu -b)
+C_{\Lambda'} D_{\Lambda'}(-v+\mu +b)] \dd v
\nonumber\\
&& +   e^{\mu b-b^2/2}(\Lambda-\Lambda')A_\Lambda A_{\Lambda'} \int_0^{\infty}D_\Lambda(v+\mu-b)D_{\Lambda'}(v+\mu-b) \dd v
\nonumber\\
&=& e^{-\mu b-b^2/2} (\Lambda-\Lambda')\int_{a-\mu-b}^{-\mu-b} [B_\Lambda D_\Lambda(v)
+C_\Lambda D_\Lambda(-v) ] [ B_{\Lambda'} D_{\Lambda'}(v)
+C_{\Lambda'} D_{\Lambda'}(-v) ] \dd v
\nonumber\\
&&+  e^{\mu b-b^2/2} (\Lambda-\Lambda')A_\Lambda A_{\Lambda'} \int_{\mu-b}^{\infty}D_\Lambda(v)D_{\Lambda'}(v) \dd v
\nonumber\\
&=&  e^{-\mu b-b^2/2}\big\{-B_\Lambda D_{\Lambda+1}(a-\mu-b)[B_{\Lambda'} D_{\Lambda'}(a-\mu-b)+C_{\Lambda'} D_{\Lambda'}(\mu-a+b)]
\nonumber\\
&&+B_{\Lambda'} D_{\Lambda'+1}(a-\mu-b)[ B_\Lambda D_{\Lambda}(a-\mu-b)+C_\Lambda D_{\Lambda}(\mu-a+b) ]
\nonumber\\
&&+C_\Lambda D_{\Lambda+1}(\mu-a+b)[ B_{\Lambda'} D_{\Lambda'}(a-\mu-b)+C_{\Lambda'} D_{\Lambda'}(\mu-a+b) ]
\nonumber\\
&&-C_{\Lambda'} D_{\Lambda'+1}(\mu-a+b)[ B_\Lambda D_{\Lambda}(a-\mu-b)+C_\Lambda D_{\Lambda}(\mu-a+b) ]\big\}.
\label{im}
\end{eqnarray}
\end{widetext}
For the last computational step we have used the properties (\ref{ig}) and
the analogous identity
\begin{equation} \label{ig1}
\frac{\dd  e^{z^2/4}D_{\nu}(z)}{\dd z}=\nu  e^{z^2/4}D_{\nu-1}(z).
\end{equation}
Indeed, if we choose for $\Lambda$ and $\Lambda'$ two different eigenvalues we obtain (bi-)orthogonality
of the eigenfunctions if the characteristic equation (\ref{ij}) is taken into account.
Furthermore dividing Eq.~(\ref{im}) on both sides by $ \Lambda-\Lambda' $ and
taking the limit $ \Lambda'\rightarrow \Lambda $ we end up with the normalization factor
\begin{widetext}
\begin{eqnarray}
 Z_\Lambda = e^{-\mu b-b^2/2} \left[ B_{\Lambda}D_{\Lambda+1}(a-\mu-b)-C_{\Lambda}D_{\Lambda+1}(\mu-a+b) \right]
 \partial_{\Lambda}\left[ B_{\Lambda}D_{\Lambda}(a-\mu-b)+C_{\Lambda}D_{\Lambda}(\mu-a+b) \right]. \label{in}
\end{eqnarray}
\end{widetext}


\end{document}